\documentclass[twocolumn,english,aps,prb,reprint,nofootinbib]{revtex4-1}
\usepackage[T1]{fontenc}
\usepackage[latin9]{inputenc}
\usepackage{textcomp,hyperref}
\usepackage{amsmath}
\usepackage{amssymb}
\usepackage{graphicx}
\usepackage{esint}

\makeatletter

\newcommand{\lyxmathsym}[1]{\ifmmode\begingroup\def\b@ld{bold}
  \text{\ifx\math@version\b@ld\bfseries\fi#1}\endgroup\else#1\fi}

\@ifundefined{textcolor}{}
{%
 \definecolor{BLACK}{gray}{0}
 \definecolor{WHITE}{gray}{1}
 \definecolor{RED}{rgb}{1,0,0}
 \definecolor{GREEN}{rgb}{0,1,0}
 \definecolor{BLUE}{rgb}{0,0,1}
 \definecolor{CYAN}{cmyk}{1,0,0,0}
 \definecolor{MAGENTA}{cmyk}{0,1,0,0}
 \definecolor{YELLOW}{cmyk}{0,0,1,0}
}

\usepackage{color}
\usepackage{babel}
\usepackage{float}
\usepackage{braket}
\usepackage{babel}
\hypersetup{colorlinks=true,citecolor=blue}
\makeatother

\newcommand{\Tm}{T$_{\text{\textbf{--}~}}$}
\newcommand{\To}{T$_{\text{0~}}$}
\newcommand{\Tp}{T$_{\text{\textbf{+}~}}$}
\newcommand{\Sx}{S$^{'~}$}

\begin{document}

\author{J. Le\'{s}nicki}

\affiliation{AGH University of Science and Technology, Faculty of Physics and
Applied Computer Science,\\
 al. Mickiewicza 30, 30-059 Kraków, Poland}

\author{B. Szafran}

\affiliation{AGH University of Science and Technology, Faculty of Physics and
Applied Computer Science,\\
 al. Mickiewicza 30, 30-059 Kraków, Poland}

\title{Spin separation and exchange for quantum dots in the Overhauser field}
\begin{abstract}
We describe the spin and charge dynamics of the system of two electrons confined within
a double quantum dot defined in a quantum wire. The spin dynamics is driven by the electron motion
in presence of the spin-orbit interaction and the randomly varying local Overhauser field due to the nuclear spins. The Schroedinger equation is solved
with the time-dependent configuration interaction method that allows for an exact description of the system
dynamics. The procedures of the spin separation, exchange and read-out by the spin to charge conversion 
all induced by the detuning variation are simulated. The 
rates of the potential variation that are necessary for the spin separation and spin to charge conversion 
in the context of the Landau-Zener transitions are determined. The average over random configurations
of the hyperfine field produce spin exchange results which qualitatively agree with the experimental data.
\end{abstract}

\maketitle

\section{Introduction}

Construction of a universal quantum gate for information processing on spins
of electrons confined in quantum dots \cite{loss} requires implementation of a controllable coupling between the spins confined in adjacent quantum dots. 
The proposed procedure 
 \cite{Burkard} employs time evolution of the system with
switching on the exchange energy \cite{harju,rontani,dassarma} $J$, defined as 
the difference of the singlet and triplet energy levels, for a short time. A nonzero exchange energy 
requires interdot tunnel coupling, and the
control can be achieved by modulation of the tunnel coupling between the dots with
either a tunable interdot barrier \cite{Burkard} or tunable potential inside one
or both quantum dots \cite{Petta05}.

The exchange energy \cite{Burkard} in the absence of spin-orbit interaction and hyperfine interaction
is isotropic \cite{Baruffa}, i.e. depends only  on  relative  orientation  of  the  spins and conserves the 
total spin in the dynamics of the few-electron systems. Variation of the spin polarization of the electron system 
is possible in presence of the spin-orbit coupling which translates the electron motion in space
to rotations of its spin as it precedes in the effective magnetic field that accompanies the spin-orbit coupling \cite{ema1,ema2,ema3,ema4,ema5,ema6,ema7}.
The spin-orbit coupling allows also for initialization of the state of the spin qubits, for separation of the
spins of moving electrons \cite{Szumniak,Pawlowski,Nitta34} in particular.

In the present paper we report on a simulation of the spin separation and spin exchange in a GaAs double quantum dot two-electron system.
We use a quasi one-dimensional model with the assumption of a strong lateral confinement and the time-dependent configuration-interaction approach
that allows for an exact account of the two-electron spin and charge dynamics.
For separation of the spins we use the texture of the internal magnetic field
arising due to a non-zero magnetic moments of atomic nuclei \cite{hanson,johnson05,bracker05}. 
The field is expressed as a local classical Overhauser field obtained by
summation of randomly oriented nuclear magnetic moments over the span of the
lateral electron wave function \cite{Osika13}. 

The coupling of the carrier spins to the changing nuclear spin field is considered the main source of dephasing and spin relaxation in III-V materials \cite{hanson}.
Here, the fluctuations of the nuclear field are used as a resource for 
 separation of the spins, which is performed via an adiabatic evolution of the system kept in the ground state when the 
tunnel coupling between the dots is quenched. The exchange interaction is then switched on for the electrons to exchange their
spins, as in Ref.~\onlinecite{Petta05}. 
Next, the procedure of the read out of the spin exchange result is implemented by
projecting the spin states on the charge configurations of the double dot. 
The spin exchange probability averaged over a number of fluctuations of the
Overhauser field reproduce the fringe pattern of the experimental results
of Ref.~\onlinecite{Petta05} as a function of the potential difference
in quantum dots (which is related to detuning~$\varepsilon$) and the spin exchange time. We discuss the rate of the potential modulation for preparation of the initial states, the effects of the potential asymmetry on the exchange interaction, and the reproducibility of the spin manipulation procedures in presence of the Rashba spin-orbit interaction 
and the Overhauser field.

\section{Theory.}

We consider a quasi one-dimensional double quantum dot defined within a GaAs quantum wire.
The single-electron Hamiltonian for the considered system reads 
\begin{equation}
	\hat{H}_{3D(1)}(t) = \frac{\hbar^2 \vec{k}^2}{2m^*} + V(\vec{r}) +
	\hat{H}_B +
	\hat{H}_{SO}
	\label{hamiltonian3d1e},
\end{equation}
with $\vec{k} = -i\nabla$, the electron effective mass in GaAs $m^* =$~0.067~$m_0$,
and $V(\vec{r})$ stands for the confinement potential. The third term
($\vec{B}_{HF}$), 
\begin{equation}
	\hat{H}_B =
	\frac{g \mu_b}{2}\vec{\sigma}\cdot (\vec{B} +
	\vec{B}_{HF}(\vec{r})),
\end{equation}
accounts for the spin Zeeman effect that includes  the external magnetic field $\vec{B}$ and
the internal  Overhauser  field due to the hyperfine interaction \cite{hanson} with the nuclear spins,
where $g = -0.44$ is the GaAs electron Land\'e factor, $\mu_b$ the Bohr magneton, and $\vec{\sigma}$
the vector of Pauli matrices.
The last component of the Hamiltonian \eqref{hamiltonian3d1e} is responsible for the Rashba-type spin-orbit coupling.
\begin{equation}
	\hat{H}_{SO} = \alpha\!\left( \sigma_z k_x - \sigma_x k_z \right).
\end{equation}
We apply  the coupling constant $\alpha =$~0.44~meV~nm \cite{DeAndrada97,Osika13}.

We assume that the system is strongly confined along the axis of the wire, so that
the electrons occupy the ground state of the lateral quantization only and
the charge dynamics involves time evolution exclusively in the axial direction. 
For the lateral single-electron wave functions we adapt
the Gaussian form,
\begin{equation}
	\Psi(x, y) = \frac{1}{\sqrt{\pi}l} \exp(-\frac{x^2 + y^2}{2 l^2}),
	\label{wave3d1exy}
\end{equation}
with $l=10$ nm. 

\begin{figure}[htbp]
	\includegraphics[width=3.4in]{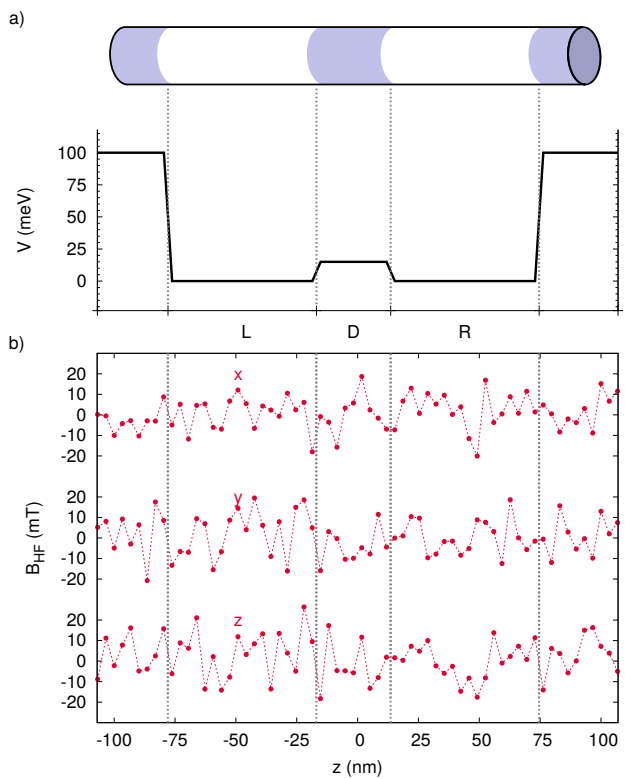}
	\caption{\label{1-scheme}
	(a) Schematic picture of fragment of a quantum wire hosting a double quantum
	dot, and its representation as a value of potential $V[z]$ along the $z$-axis. 
	(b) A sample of the effective HF magnetic (Overhauser) field generated at random with the approach of Ref.~\onlinecite{Osika13}.
	Each vector component ($x$, $y$, $z$) was plotted separately.
	} \label{schemat}
\end{figure}

The Hamiltonian \eqref{hamiltonian3d1e} neglects the orbital effects of the
external field which is justified by the assumption of a strong lateral confinement.
Based on this assumption we consider  the electron wave functions frozen in the form given by Eq. \ref{wave3d1exy} which allows
us to introduce a quasi-one-dimensional version of the energy
operator (\ref{hamiltonian3d1e}) obtained by integrating the Hamiltonian
$\langle\Psi|\hat{H}_{3D(1)}|\Psi\rangle$ over $x$ and $y$ dimensions,
which yields:
\begin{equation}
	\hat{H}_{(1)} =
	-\frac{\hbar^2}{2m^*}\frac{\partial{}^2}{\partial{}z^2} +
	V(z) +
	\frac{g \mu_b}{2}\vec{\sigma}\cdot (\vec{B} +
	\vec{B}_{HF}(z)) +
	i\alpha\sigma_x \frac{\partial{}}{\partial{}z},
	\label{hamiltonian1d1e}
\end{equation}
where $V(z)$ is the confinement potential along the axis of the wire
 [Fig.~\hyperref[schemat]{\ref*{schemat}(a)}],
 with two quantum dots of length $L$ and $R$,
separated by a barrier of length $D$. Two barrier regions fill the remainder of the computational box of length 213.6 nm.
We took $L = R =$~61.02~nm and $D =$~30.51~nm
for all the presented results.
The potential is taken 0 inside the right quantum dot, 15~meV within the 
central barrier, and 100~meV in the outer barriers [Fig.~\hyperref[schemat]{\ref*{schemat}(a)}].
The potential in the left quantum dot is set equal to $\Delta V$ that is varied in the simulation. 

In most of the calculations the external magnetic field is taken  $|\vec{B}| =$~100~mT as in the experiment~\onlinecite{Petta05}. For this value, the nuclear spins separate from the electron spins
in the sense that the electron-nuclear spin flips \cite{sepa1,sepa2,hanson} are forbidden by the energy conservation.
We approximate the hyperfine field by a local magnetic field of a random orientation which is considered constant
during the time evolution of the electron spin. The latter assumption is justified by a long fluctuation time of the nuclear
field which is of the order of 10 to 100 $\mu$s \cite{hanson}. The fluctuation of the hyperfine field in the experimental conditions
is accounted for by averaging the results of the spin dynamics -- which is also used in the experimental data processing \cite{Petta05}.

The procedure for derivation of the field $\vec{B}_{HF}(z)$ is adapted from Ref.~\onlinecite{Osika13} 
with a classical vector of the effective magnetic field of length 5T generated at random orientation at each ion of the crystal followed by the averaging of all the local nuclear vectors with the probability density corresponding to the Gaussian lateral wave functions (\ref{wave3d1exy}).
 A sample of the generated fields is plotted in Fig.~\hyperref[schemat]{\ref*{schemat}(b)}.
The values in Fig.~\hyperref[schemat]{\ref*{schemat}(b)} are given for each grid point in the finite difference approach that we employ for determination of the single-electron eigenstates.
The effective HF field as seen by the electron spin is further reduced to a few milliteslas \cite{johnson05,bracker05} by averaging over the probability density 
along the axis of the quantum dot.  

The integration of the two-electron stationary Hamiltonian over the lateral degrees of freedom produces the operator
\begin{equation}
	\hat{H}_{(2)}^{(0)} = \sum_{i=1}^2{\hat{H}_{(1)}(\sigma_i, z_i)} +
	\frac{\sqrt{\pi/2}~e^2}{4\pi \varepsilon_0 \varepsilon l}~\mathrm{erfcx}\!\left(
	\frac{|z_1 - z_2|}{\sqrt{2}l} \right),
	\label{hamiltonian1d2e}
\end{equation}
where the last term is the electron-electron interaction potential that results from the integration
of the Coulomb interaction with the lateral wave function\cite{pbed},  $\varepsilon=12.9$ is the dielectric constant,
and $\mathrm{erfcx}$ is the scaled complementary error function.

The two-electron Hamiltonian \eqref{hamiltonian1d2e} is diagonalized by the configuration interaction method with the basis of Slater determinants $\chi_m$
of antisymmetrized products of the single-electron eigenfunctions of operator \eqref{hamiltonian1d1e},
\begin{equation}
	\psi_n(\sigma_1, \sigma_2, z_1, z_2) =
	\sum_{m=1}^{M} v_{nm} \chi_m(\sigma_1, \sigma_2, z_1, z_2).
	\label{cfgs}
\end{equation}
We used at least 120 Slater determinants for the basis \eqref{cfgs} to obtain the eigenstates $\psi_n$
and the energies of the two-electron Hamiltonian. 

For simulation of the system dynamics 
we separate the external potential variation from the time-independent Hamiltonian,
\begin{equation}
	\hat{H}_{(2)}(t) = \hat{H}_{(2)}^{(0)} + \hat{W}_{(2)}(t)
	\label{perturbed-hamiltonian}.
\end{equation}
The simulated potential variation amounts in changing the $\Delta{}V$ (additional left quantum dot potential)
\begin{equation}
	\hat{W}_{(2)}(t) = \Delta{}V(t) \cdot (l(z_1) + l(z_2)),
	\label{perturbation}
\end{equation}
where $l(z)$ is equal 1 in the left quantum dot and 0 elsewhere.

We solve the time-dependent Schr\"odinger equation 
\begin{equation}
	i\hbar\frac{\partial}{\partial{}t}\psi(t)=
	\hat{H}_{(2)}^{(0)}\psi(t) +
	\hat{W}_{(2)}(t)\psi(t),
	\label{perturbed-time-dep-schr-eqn}
\end{equation}
in the  basis of $\hat{H}_{(2)}^{(0)}$ eigenstates $\psi_n$ (\ref{cfgs}):
\begin{equation}
	\psi(t) = \sum_{n=1}^{N}{a_n(t) \psi_n(\sigma_1, \sigma_2, z_1, z_2)\exp{(-iE_n t/\hbar)}}.
\end{equation}
This form of the wave function when plugged into the Schr\"odinger equation 
produces a set of differential equations for the $a_k$ coefficients,
\begin{equation}
	\dot{a}_k(t) =
	-\frac{i}{\hbar} \sum_{n=1}^{N}
	a_n(t) \langle\psi_k| \hat{W}_{(2)}(t) |\psi_n\rangle e^{i (E_k - E_n) t / \hbar},
\end{equation}
that we solve using the Crank-Nicolson scheme,
for which the subsequent steps of the wave function 
are given by  solution of an algebraic linear system of equations,
\begin{equation}
	\left[ I - \frac{1}{2} W(t + \Delta{}t) \Delta{}t \right]
	\vec{a}(t + \Delta{}t)
	= \left[ I + \frac{1}{2} W(t) \Delta{}t \right] \vec{a}(t),
\end{equation}
with $I$ that stands for the identity matrix, and $W(t)$ that stands for a
matrix with elements
\begin{equation}
	W_{k,n}(t) = 
	-\frac{i}{\hbar} \Delta{}V(t) 
	\langle\psi_k| l(z_1) + l(z_2) |\psi_n\rangle e^{i (E_k - E_n) t / \hbar}.
	\label{w-elements}
\end{equation}

\section{Results}

\subsection{Two-electron eigenstates}

\begin{figure}[htbp]
	\includegraphics[width=3.4in]{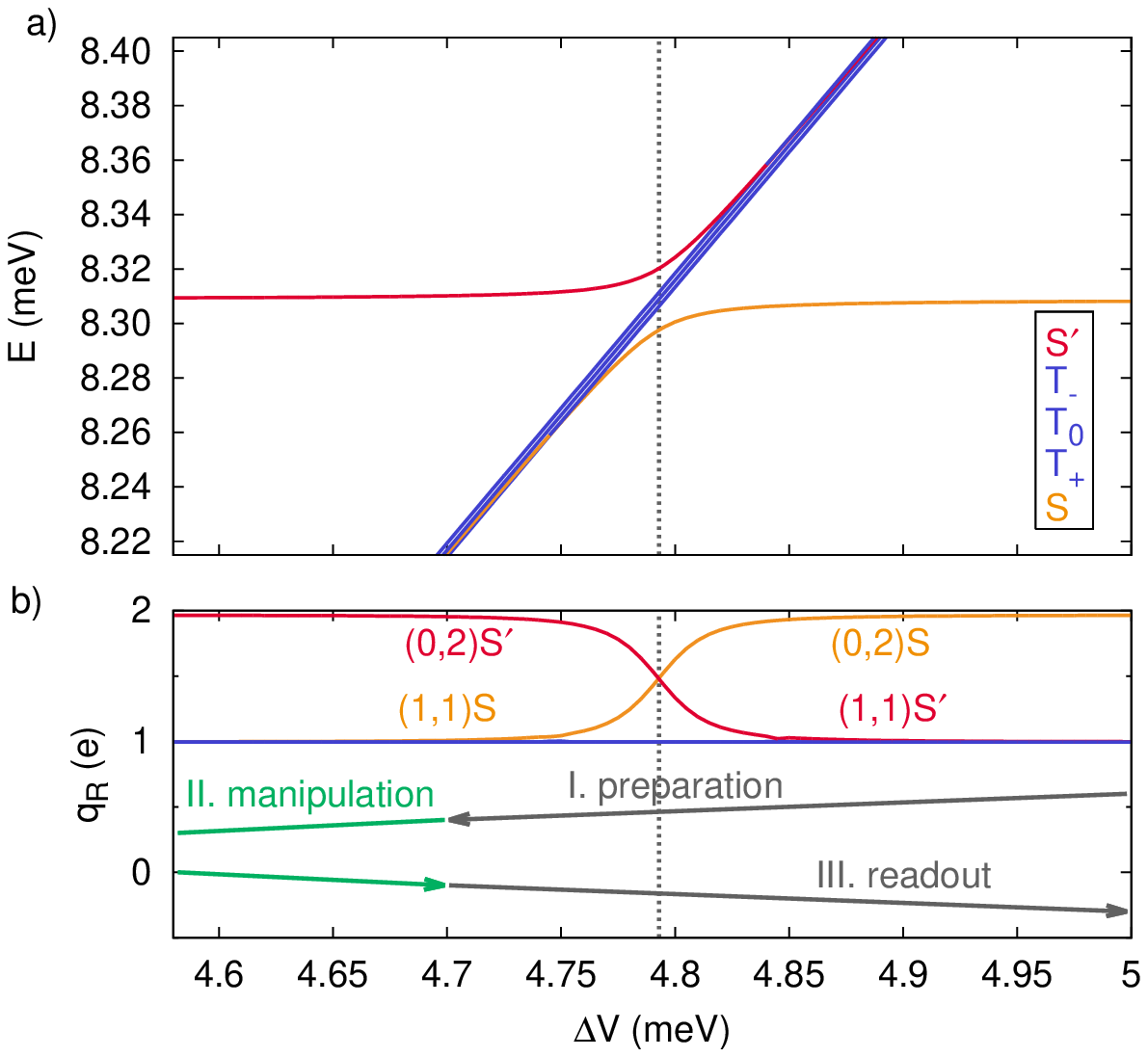}
	\includegraphics[width=3.4in]{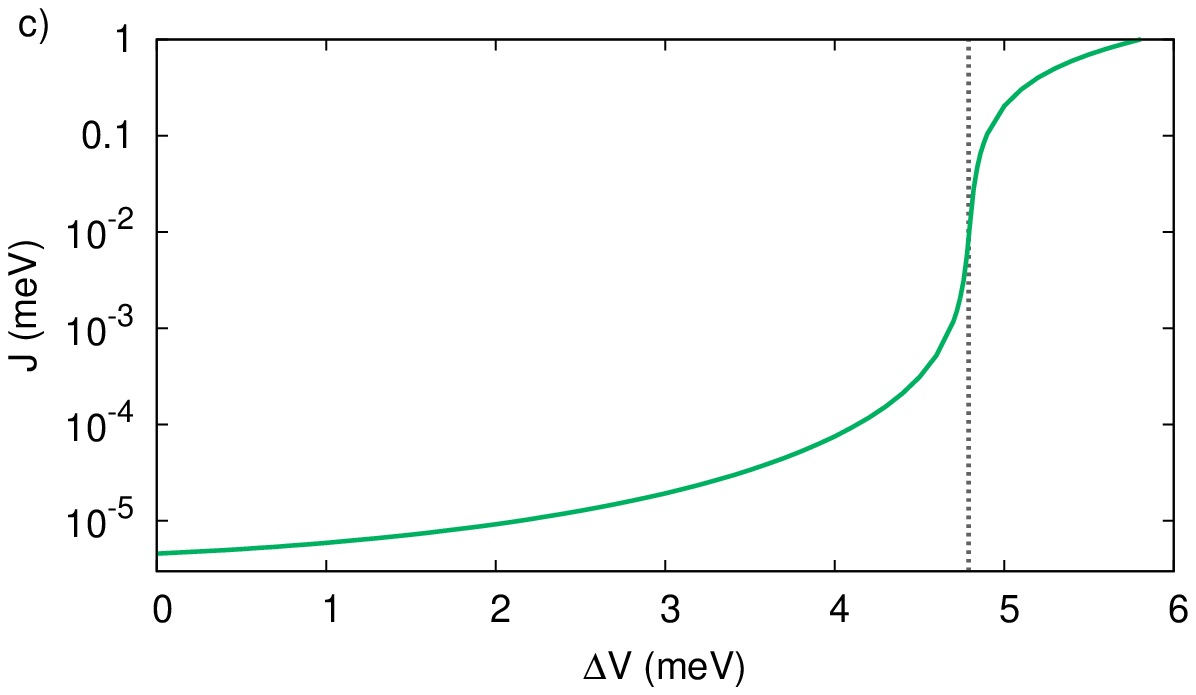}
\caption{
	(a) Energies $E$ and (b) electron charge $q_R$ stored by the right quantum dot,
	 as functions of the potential difference $\Delta{}V$. The sequence of $\Delta{}V$ changes
	is presented in (b).  The system preparation amounts to an adiabatic charge transfer from (0,2)S state 
         to the state with separated electrons (1,1)S. 
          The manipulation of the state is carried at $\Delta V$ below the avoided crossing. 
          For the readout of the spin state after the manipulation stage the system is ramped back to the voltages
        for which (0,2)S is the ground state. 
	An axial magnetic field of $B_z=100$ mT is applied here and below, unless stated otherwise.
        (c) The exchange energy $J$ defined as the energy difference between $T_0$ and the lowest-energy singlet~S. 
           The center of the avoided crossing is marked by the vertical dotted line.
	}\label{spe}
\end{figure}

The lowest two-electron energy levels  are plotted in Fig. \hyperref[spe]{\ref*{spe}(a)} as functions of the difference of potentials in left and right quantum dot.
The charge localized in the right dot for the corresponding energy levels is plotted in Fig. \hyperref[spe]{\ref*{spe}(b)}.  
The splitting of the triplet energy levels in Fig. \hyperref[spe]{\ref*{spe}(a)} is due to the axial magnetic field 
 set to $B_z=100$ mT.
The spectrum in Fig. \hyperref[spe]{\ref*{spe}(a)} contains the triplet states with a single electron per quantum dot [charge configuration denoted by (1,1)] as well as the singlet states with separated electrons (1,1)S, and both electrons localized in the right quantum dot (0,2)S (\Sx in the following denotes the higher-energy singlet state). 
Figure \hyperref[spe]{\ref*{spe}(a)} shows an avoided crossing of the singlets near $\Delta V=4.79$ meV. The (0,2)S energy 
level does not react to the potential variation in the left quantum dot with $\Delta V$, hence its weak dependence on the potential variation outside the avoided crossing.

\begin{figure}[htbp]
	\includegraphics[width=3.4in]{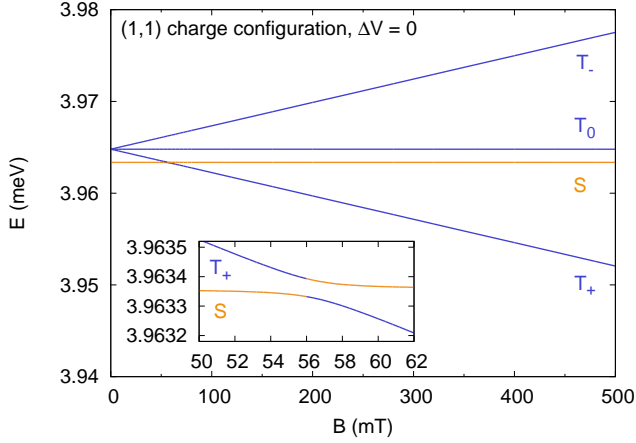}
	\caption{\label{1-spect-varmagn}
	Energies of the lowest four states of the system as a function of external magnetic
	field $B$. Inset shows enlarged anticrossing of states S and \Tp\!. Curves are
	colored according to symmetry (singlet/triplet) and not to the actual order of
	states (which changes due to crossings). \label{mf}}
\end{figure}

The magnetic field dependence of the energy levels with separated electrons --
obtained for symmetric confining potential (i.e. for $\Delta V=0$) --
is depicted in Fig. \ref{mf}. 
The  narrow  avoided crossing between the S and \Tp energy levels near 56 mT is induced by the spin mixing factors of the HF field and the SO coupling.

In the absence of the HF field and the SO coupling the wave functions of the two-electron eigenstates are separable into
the spatial and spin components, in particular, for the states with zero total
spin projection:
\begin{equation}
\Psi_\mathrm{S}=\psi_\mathrm{S}(z_1,z_2)\frac{1}{\sqrt{2}}\left(\chi_\uparrow(\sigma_1)\chi_\downarrow(\sigma_2)- \chi_\uparrow(\sigma_2)\chi_\downarrow(\sigma_1)\right), \label{str}
\end{equation}
and
\begin{equation}
\Psi_{\mathrm{T}_0}=\psi_{\mathrm{T}_0}(z_1,z_2)\frac{1}{\sqrt{2}}\left(\chi_\uparrow(\sigma_1)\chi_\downarrow(\sigma_2)+ \chi_\uparrow(\sigma_2)\chi_\downarrow(\sigma_1)\right), \label{ttr}
\end{equation}
where $\chi$ are the spin eigenstates. For the states \eqref{str} and \eqref{ttr} the spin-up and spin-down densities are equal in each point in space, 
so that the dots store zero average spin.
The spatial wave functions for states \eqref{str} and \eqref{ttr}  can in the first approximation  be expressed by the single-dot $\phi_l$ and $\phi_r$ ground-state orbitals, localized
in the left and right dots, respectively
\begin{equation}
\psi_\mathrm{S}(z_1,z_2)=\frac{1}{\sqrt{2}} \left(\phi_l(z_1)\phi_r(z_2)+\phi_r(z_1)\phi_l(z_2)\right),
\end{equation}
\begin{equation}
\psi_{\mathrm{T}_0}(z_1,z_2)=\frac{1}{\sqrt{2}} \left(\phi_l(z_1)\phi_r(z_2)-\phi_r(z_1)\phi_l(z_2)\right).
\end{equation}
In the weak interdot tunneling regime the S and \To states are nearly degenerate and can be mixed by either
the spatial variation of the effective Land\'e $g$ factor \cite{NadjPerge,schroer11,frolov12}, the spin-orbit coupling \cite{Nowak15}, or the HF field \cite{Petta05}. 
For the maximal mixing case one obtains the spin separation over the dots \begin{eqnarray}&& \Psi_{\uparrow\downarrow}=\frac{1}{\sqrt{2}}\left( \Psi_\mathrm{S}+\Psi_{\mathrm{T}_0} \right)= \\ && \nonumber \frac{1}{\sqrt{2}}\left(\phi_l(z_1)\chi_\uparrow(\sigma_1)\phi_r(z_2)\chi_\downarrow(\sigma_2)-\phi_l(z_2)\chi_\uparrow(\sigma_2)\phi_r(z_1)\chi_\downarrow(\sigma_1)\right)\label{ud}\end{eqnarray}
with the left (right) dot storing the spin-up (spin-down) density
and a state with interchanged spins
\begin{eqnarray}&& \Psi_{\downarrow\uparrow}=\frac{1}{\sqrt{2}}\left( \Psi_\mathrm{S}-\Psi_{\mathrm{T}_0} \right)=\\&&\nonumber \frac{1}{\sqrt{2}}\left(\phi_r(z_1)\chi_\uparrow(\sigma_1)\phi_l(z_2)\chi_\downarrow(\sigma_2)-\phi_r(z_2)\chi_\uparrow(\sigma_2)\phi_l(z_1)\chi_\downarrow(\sigma_1)\right).\label{du}\end{eqnarray}

\begin{figure}[htbp]
	\includegraphics[width=3.4in]{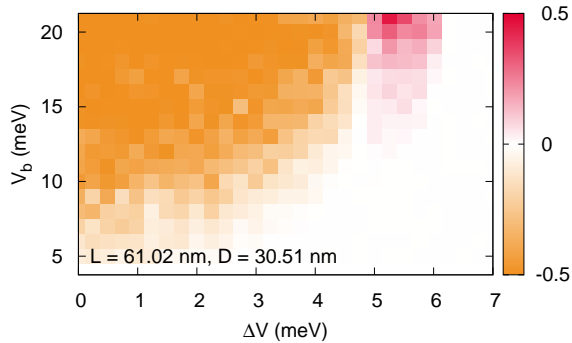}
\caption{
	Averaged absolute values of spin $s_R$ in the right quantum dot in the second excited state, for different values of dot potential difference $\Delta{}V$ and barrier
	potential $V_b$.
	Values are taken with negative sign (orange) if the separation involves S and \To
	states, and with positive sign (red) when  \To and \Sx are involved.
	} \label{sepszpi}
\end{figure}

The spin-orbit coupling alone can separate the spins over the dots in the external field but only provided that the double dot system is strongly asymmetric,
with one dot larger by a factor of three than the other \cite{Nowak15}. 
We find that the spin separation by the HF field occurs also for quantum dots of the same size.
  The  average spin in the right dot calculated for the second excited state 
 (\To\!) is displayed in Fig. \ref{sepszpi}.
 The average was taken over five random HF field distributions.
 Naturally, for each random distribution the average spin in the left and right quantum dots is different.
  Note, that the experiment \cite{Petta05} also applies
averaging the results of the spin evolution over many runs -- for which the HF field varies. 
Figure \ref{sepszpi} shows that for a high energy barrier the spin in the right dot is close to $|\langle S_z \rangle | =\frac{\hbar}{2}$,
i.e. the spin separation is a typical result. We find that the spin configuration in the second excited state is then either $\uparrow\downarrow$ or $\downarrow\uparrow$,
depending on the specific HF field distribution,
the sign of the spin was hence neglected, instead, the values were taken as
negative (orange) for the cases where the second state participating in the spin
separation is the ground singlet S.
In nanowire double quantum dots a substantial variation of the $g$ factors in both the dots
has been found \cite{schroer11,frolov12,NadjPerge}. The variation should fix the orientation of the spins in the Hamiltonian eigenstates in the absence of
the exchange interaction. 
The spin separation in the Hamiltonian eigenstates is very rarely  obtained for the $\Delta V$ values which correspond
to the avoided crossing between the (1,1) and (0,2) singlets, near $\Delta V=5$ meV. In this energy range the exchange energy is strong
and prevails over the HF field fluctuations.
Figure \ref{sepszpi} shows that the closer we are to this value, the larger the interdot barrier $V_b$ needs to be in order to induce the spin separation. 
The spin separation in the weak coupling regime is crucial for the charge and spin dynamics to be discussed below.

\begin{figure*}[htbp]
\includegraphics[width=7in]{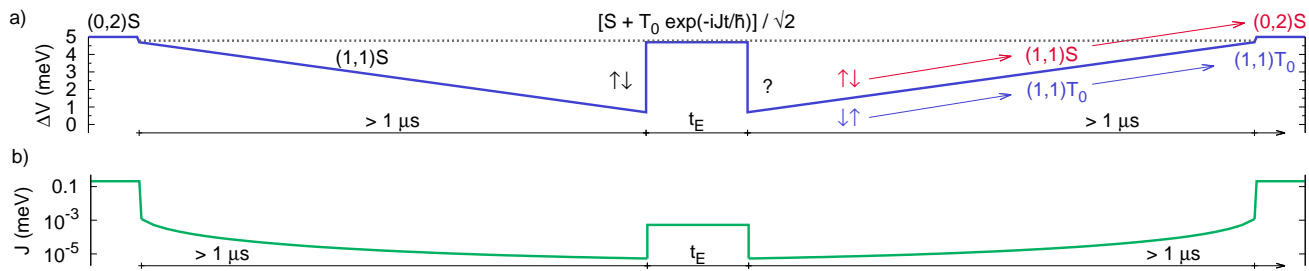}
\caption{ (a) Variation of the potential asymmetry in time with the targeted evolution of the two-electron state explained schematically. (b) The exchange energy defined as the energy difference between the triplet \To state and the lowest energy singlet for the potential asymmetry in (a).
	} \label{swap}
\end{figure*}

\subsection{The spin separation and exchange sequence}

The spin separation and exchange procedure that is simulated below is adapted from the experiment of Ref.~\onlinecite{Petta05} and depicted in Fig. \ref{swap}.
The procedure starts with a strong potential difference $\Delta V$ with two electron ground state singlet localized in the right quantum dot S(0,2). 
A slight change of $\Delta V$ is applied to pass across the singlets avoided crossing of Fig. \ref{spe} with the evolution time that is fast on the time scale of the hyperfine field spin flipping but slow on the time scale defined by 
the exchange  interaction,  in order to change the charge occupation of the dots but keep the singlet spin state.
Next, the
symmetry of the confinement potential is restored slowly on the hyperfine interaction time scale, which -- as we show below  -- in presence of the HF field leads to the appearance 
of the state with definite spin orientation in each of the dots. Next, large asymmetry of the potential is reintroduced for duration of $t_E$.
 The asymmetry of the confinement potential
enhances the exchange interaction \cite{Szafrana} and produces the spin flips between the dots as a result of time evolution of the superposition of (1,1)S and (1,1)\To states.
After $t_E$ the exchange energy is first rapidly quenched
and then the potential is adiabatically changed towards the initial state. The right dot is occupied by two electrons with a maximal probability provided that an even number
of spin-flips was performed during the spin exchange time $t_E$. In the subsections to follow we first explain the electron structure of the eigenstates, next we move to the description of the system
initialization, readout by the spin to charge conversion, and the interdot spin exchange.

\subsection{Time evolution: Spin-dependent charge dynamics.}

\begin{figure*}[htbp]
	\includegraphics[width=7in]{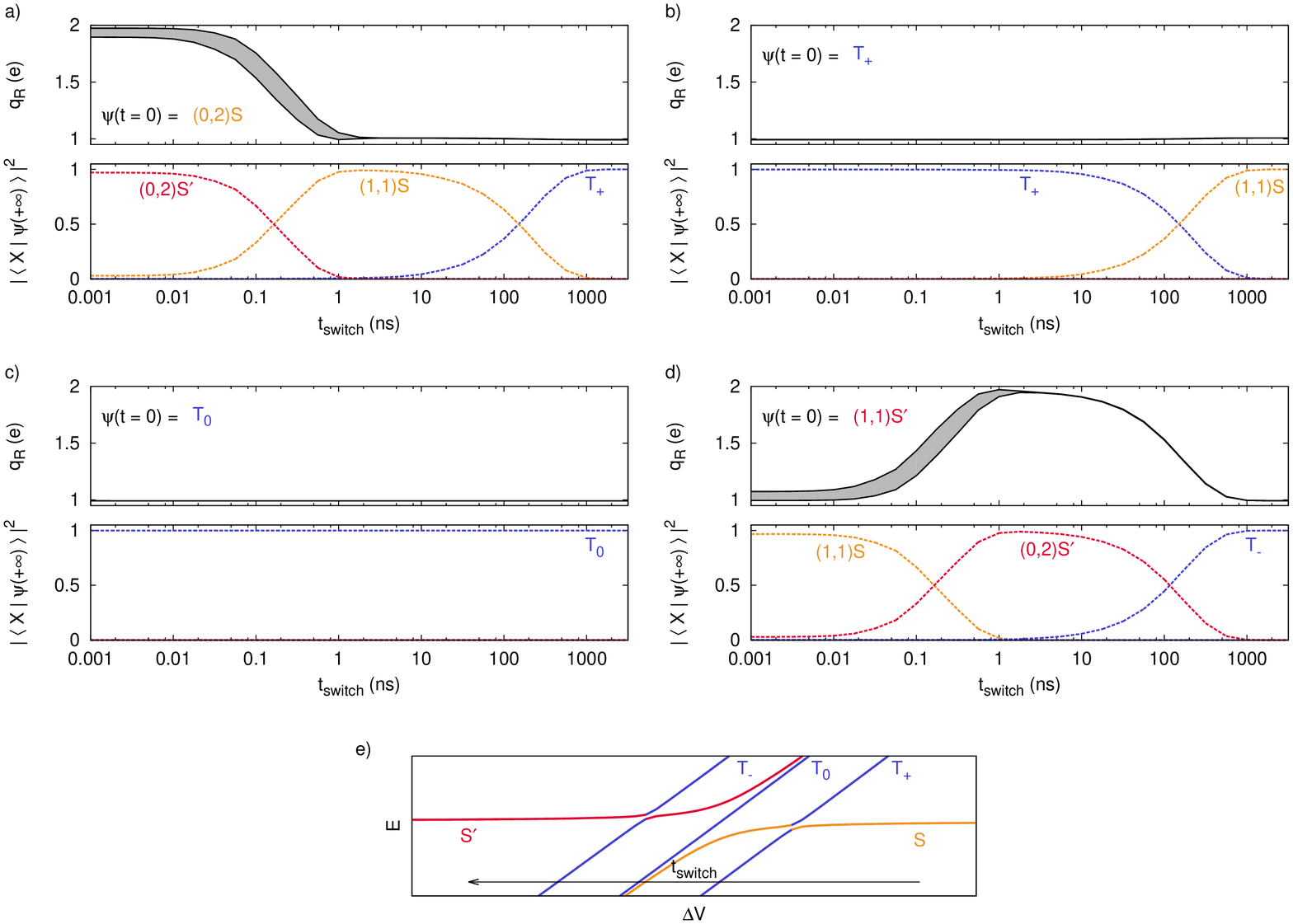}
	\caption{
	Final absolute values of charge $q_R$ in the right quantum dot (upper panels)
	and squared moduli of projections of final state $\psi(+\infty)$ onto
	the final hamiltonian basis $X \in \{$S, \Tp\!, \To\!, \Tm\!, \Sx\!\} (lower
	panels), as a function of switching time $t_{switch}$, if the system was initially
	in the (a) (0,2)S (b) \Tp (c) \To and (d) (1,1)\Sx state. Potential difference $\Delta{}V$
	was switched from 5.0~meV to 4.7~meV (state preparation). The gray areas in the upper panels of (a) and (d) indicate
        the range within which the charge in the right dot changes in the final state, which is a superposition of two final Hamiltonian eigenstates
when the switching time is short.
        (e) The schematics of the spectrum and the switching direction.
	} \label{final}
\end{figure*}

The sequence  that is  simulated  (see Fig. \hyperref[spe]{\ref*{spe}(b)} and Fig. \ref{swap})  starts as in the experiment \cite{Petta05} by 
the two-electron singlet with both electrons in the right dot (0,2)S. 
In this subsection we deal with the charge separation that is achieved by the small drop of $\Delta{}V$ from $\Delta V=5$ meV to $\Delta V=4.7$ meV that is visible at the beginning of the sequence in Fig. \hyperref[swap]{\ref*{swap}(a)}.
The drop takes the system  across the avoided crossing of the singlets in Fig. \hyperref[spe]{\ref*{spe}(a)}.
The initial states are taken as eigenstates of the stationary Hamiltonian for $\Delta V=5$ meV.
The final state of the time evolution is plotted in Fig. \hyperref[final]{\ref*{final}(a-d)} as a function of the switching time
for the first four eigenstates of the Hamiltonian for $\Delta V=5$~meV, and in
terms of the eigenstates of the Hamiltonian for $\Delta V=4.7$ meV (lower panels).
Additionally, the resultant value of charge in the right dot (upper panels)
is provided to emphasize states having (0,2) occupation.
The rate of the changes can be compared with two time scales: the one given by the energy splitting between the singlets, which at the center
of the avoided crossing [see Fig. \hyperref[spe]{\ref*{spe}(c)}] is $\Delta S\simeq 0.02$ meV that corresponds to $\tau_E\simeq \frac{\hbar}{2 \Delta{}S}=16.5$ ps,
and the one given by the Zeeman splitting in the HF nuclear field of the singlet and \Tp state, $\Delta E_Z= g\mu_B B_{HF}'$, where $B_{HF}'$ is given by averaging the nuclear magnetic field of Fig. \hyperref[schemat]{\ref*{schemat}(b)} 
with the wave function along the $z$ coordinate, typically $B_{HF}'=1$ mT, for which $\tau_{HF}\simeq \frac{\hbar}{2 \Delta E_z}=12.9$ ns.

For the (0,2)S state as the initial state [Fig. \hyperref[final]{\ref*{final}(a)}] the final one is (0,2)\Sx when
the switching is fast on the $\tau_S$ scale. The transition has then the Landau-Zener character \cite{lz1,lz2,lz3,lz4}
 The charge occupation of the dots is left unchanged for the nonadiabatic abrupt switching.
To be more precise, the abrupt potential change leaves a small admixture of (1,1)S state to (0,2)\Sx - which
produces  oscillations of the charge localized in the left dot in the limits that are marked in the upper panel of Fig. \hyperref[final]{\ref*{final}(a)} with
the gray area.   For an adiabatic  switching time comparable with $\tau_S$  the final state is the spin separated singlet (1,1)S, while
for a very slow switching time -- comparable with $\tau_{HF}$ the evolution is adiabatic on the HF coupling time scale and the destination state is the \Tp ground state	 [see Fig. \hyperref[final]{\ref*{final}(e)}]. Summarizing, Fig. \hyperref[final]{\ref*{final}(a)} indicates a sequence of
 transitions across the two avoided crossings that involves both the (1,1) and (0,2) singlets as well as the \Tp triplet.
The \Tp -- S avoided crossing is much tighter than the (1,1)S and (0,2)S one, and thus it requires a much slower potential variation 
for the electron to pass from (0,2)S to the \Tp ground state. 
Conversely, for the initial state [Fig. \hyperref[final]{\ref*{final}(b)}] set at \Tp -- an excited state at $\Delta V=5$ meV -- a very slow switching time $\simeq \tau_{HF}$
is required to keep the electron in the excited state [(0,2)S for lower $\Delta V$], otherwise the  time evolution ends at \Tp triplet.
 \To state does not enter any avoided crossing. For the initial state set at \To one stays in the \To independent of the switching time [Fig. \hyperref[final]{\ref*{final}(c)}].
Finally, for (1,1)\Sx in the initial state  the final one is (1,1)S for a fast switching (much shorter than $\tau_S$), (0,2)\Sx for a longer switching time (between $\tau_S$ and $\tau_{HF}$), and
\Tm for an extremely slow ($\simeq \tau_{HF}$) switching. Therefore, both singlets in the initial state evolve to spin-polarized triplets in the limit of slow potential variation.

\begin{figure*}[htbp]
	\includegraphics[width=7in]{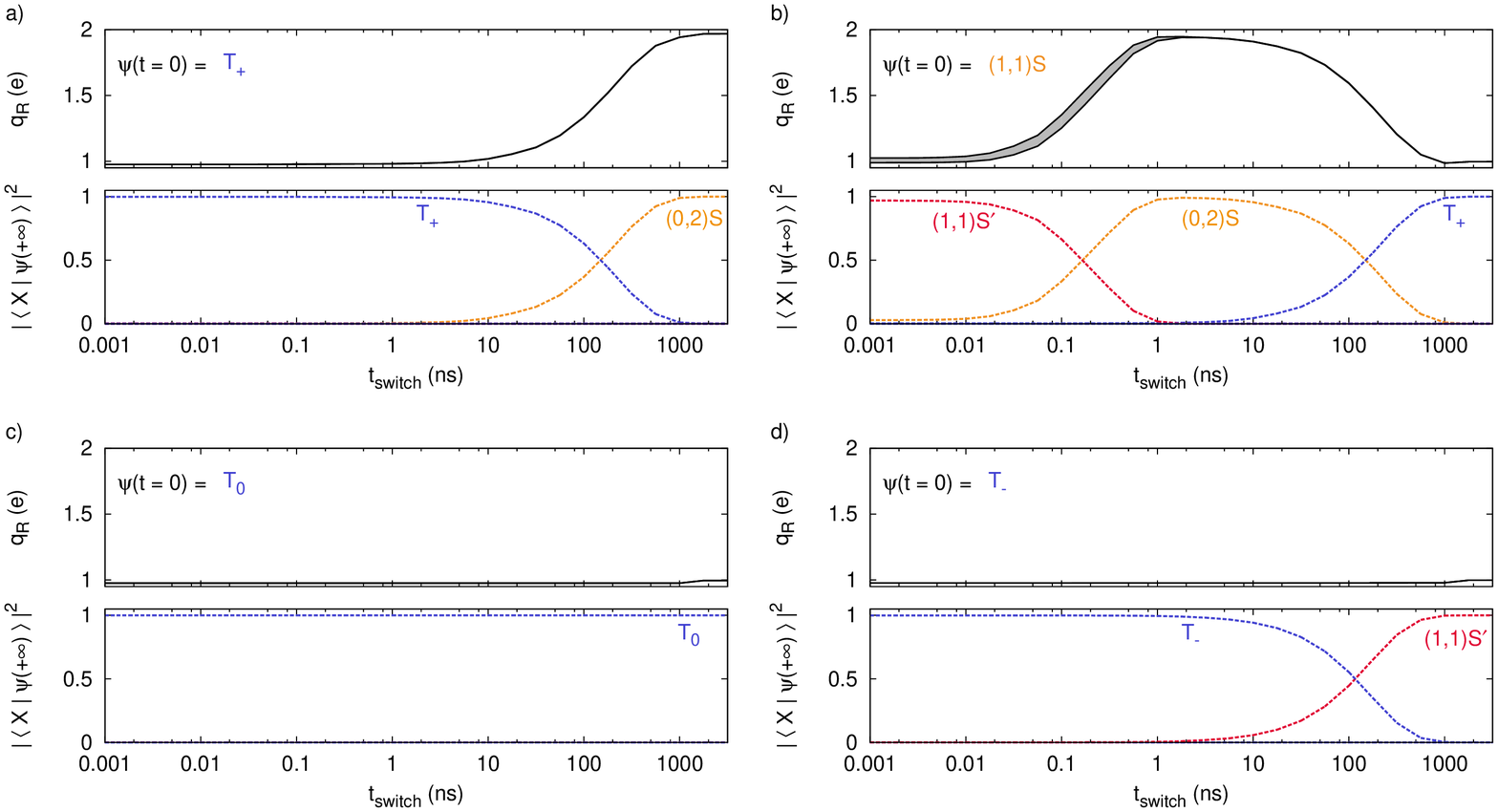}
	\caption{\label{1-e}
	Final absolute values of charge $q_R$ in the right quantum dot (upper panels)
	and squared moduli of projections of final state $\psi(+\infty)$ onto
	the final hamiltonian basis $X \in \{$S, \Tp\!, \To\!, \Tm\!, \Sx\!\} (lower
	panels), as a function of switching time $t_{switch}$, if the system was initially
	in the (a) \Tp (b) (1,1)S (c) \To and (d) \Tm state. Potential difference $\Delta{}V$
	was switched back from 4.7~meV to 5.0~meV (state readout). \label{back}
	}
\end{figure*}

The time evolution presented in Fig. \ref{final} with $\Delta V$ varied from 5 meV to 4.7 meV corresponds to the initial state preparation in the experimental sequence \cite{Petta05}.
The reverse potential variation  is used \cite{Petta05} in the spin state detection by the spin-charge conversion.
In the experiment \cite{Petta05} the charge detection of the right quantum dot is used for
determination of the result of the spin dynamics.
 The results of the simulation for the potential variation from $\Delta V=4.7$ meV to $\Delta V=5$ meV -- the small and abrupt rise of the potential in the left dot at the end of the sequence of Fig. \ref{swap}(a),
are given in Fig. \ref{back}. The initial state is set as one of the eigenstates of the Hamiltonian for $\Delta V=4.7$ meV
with the final states projected onto the Hamiltonian eigenstates for $\Delta V=5$ meV. Figure \ref{back} shows that the (1,1)S singlet evolves to (0,2)S only provided that the switching
time is longer than $\simeq 0.1$ ns but not longer than $\simeq 200$ ns. 
For a slower switching the system evolves to $T+$. A slower switching time for \Tp in the initial state produces (0,2)S in the final state [Fig. \ref{back}(a)]. 
However, in order to produce a detectable increase of the charge in the right dot above a single electron charge, the switching time needs to exceed 10 ns. 
For \Tm in the initial state a slow switching produces the spin singlet but with electrons separated over the dots [Fig. \ref{back}(d)].

\begin{figure}[htbp]
	\includegraphics[width=3.4in]{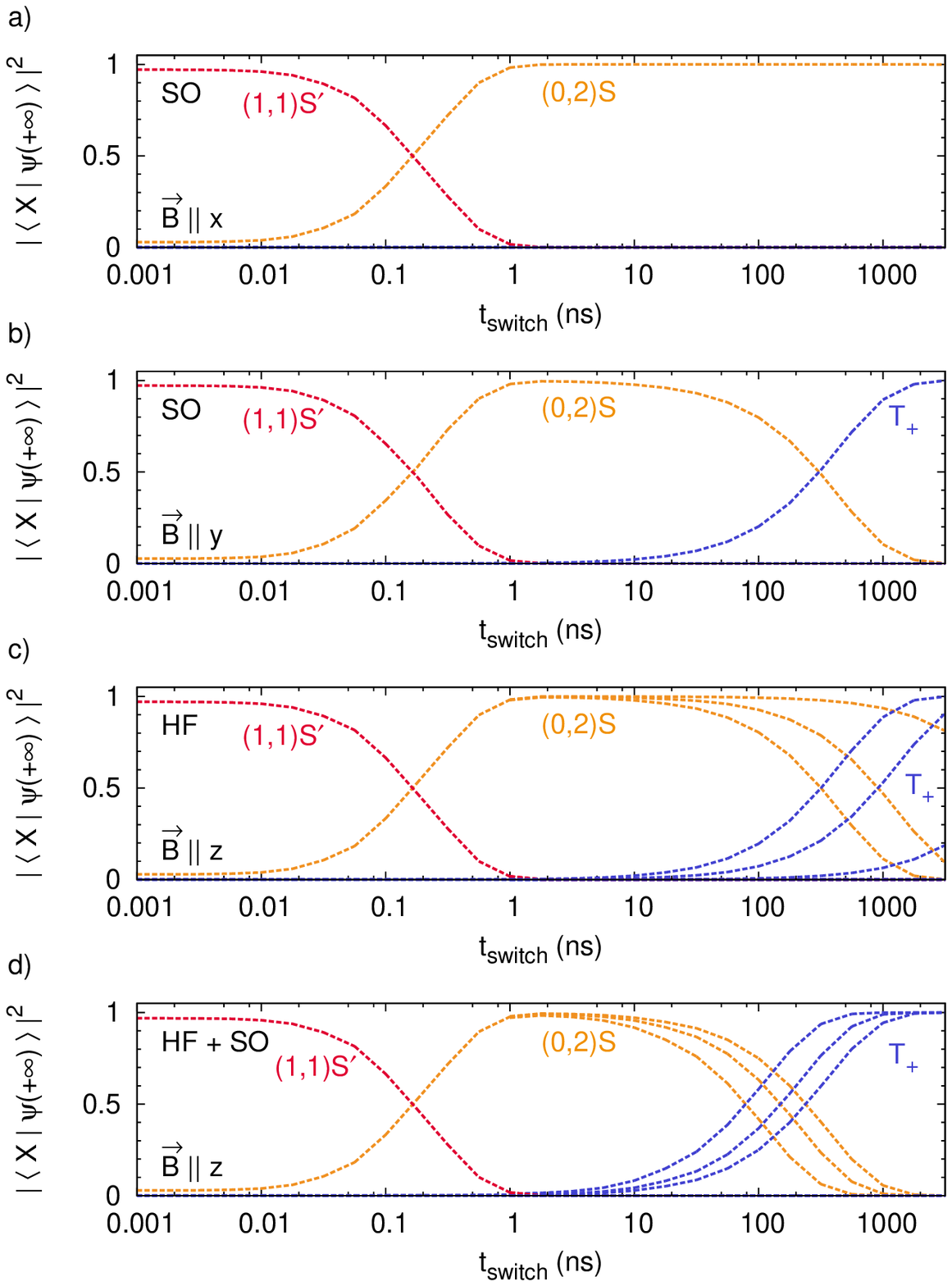}
	\caption{Same as Fig. \ref{back} for varied conditions of the simulation.  The HF field is switched off (a,b) and  the magnetic field oriented parallel to the $x$ (a) and $y$ (b) directions. The SO interaction is switched off in (c). In (d) both interactions are present. Three random configurations of the HF field were considered in (c,d). 
     } \label{ciach}
\end{figure}

First two panels of Figure \ref{ciach} shows the time evolution for the read-out sequence with the HF field switched off and the orientation of the external field 
changed to parallel to the $x$ [Fig. \ref{ciach}(a)] and $y$ directions [Fig. \ref{ciach}(b)]. For the $x$ direction of the external field the spin of the two-electron state remains unchanged.
In this case the effective SO magnetic field and the external field are aligned, so that the electron motion only changes the Zeeman splitting energy and 
no precession of the spin is present. For the external field oriented parallel to the $y$ axis the precession reappears since the external and the effective fields are no longer aligned. In presence of the HF interaction - the intrinsic anisotropy of the SO interaction is masked by the nuclear field, and the results remain quantitatively similar 
for varied external field orientation. However, the switching times differ within a certain range from one HF field configuration another. The results for three random 
configurations of the HF field are given in Fig. \ref{ciach}(c,d) with (d) or without (c) the SO interaction. The transition between the singlets
-- which are spin conserving -- ignore the details of the HF field, however the transition to the spin polarized triplet does depend on the
random HF field. Without the SO interaction the S-\Tp switching times differ by two orders of magnitude. The presence of the SO coupling 
reduces this variation range significantly - to a single order of the magnitude only.

\subsection{Time evolution: Spin-separation} 

\begin{figure}[htbp]
	\includegraphics[width=3.4in]{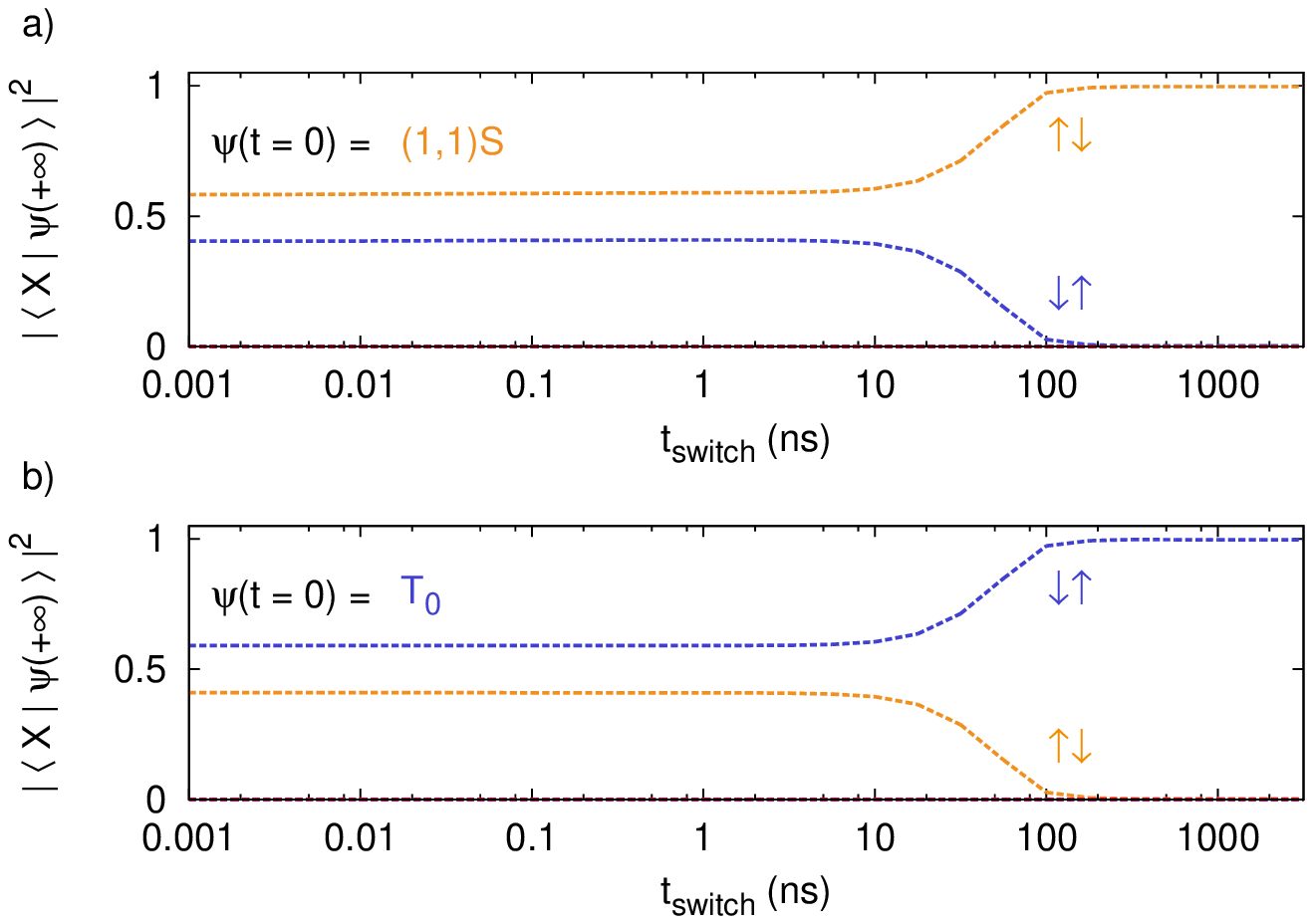}
	\caption{\label{1-a}
	Squared moduli of projections of final state $\psi(+\infty)$ onto
	the final hamiltonian states with separated spins $X \in \{\uparrow\downarrow,~
	\downarrow\uparrow\}$, as a function of switching time $t_{switch}$, if
	the system was initially in the (a) (1,1)S or (b) \To state.
	Potential difference $\Delta{}V$ was switched from 4.7~meV to 0.7~meV.
	\label{sepspi}
	}
\end{figure}

According to the preceding subsection the single-dot (0,2)S singlet can be transformed into the state with separated carriers (1,1)S 
provided that the switching time across the avoided crossing is of the order of 1 to 10 ns.  For $\Delta V=4.7$ meV -- considered above
the spins are generally not polarized within the separate quantum dots [see Fig. \ref{sepszpi}].
The spin separation in the system can then be induced by an adiabatic variation of the potential from $\Delta V$=4.7 to 0.7 meV -- 
see the part of the sequence with the slow potential difference drop in Fig. \ref{swap}(a).
The results for the system evolution are depicted in Fig. \ref{sepspi} for a typical random HF field. Figure  \ref{sepspi} shows the projections of the final states in the
basis of the destination Hamiltonian eigenstates. For the switching time that exceeds 100 ns the final states are the Hamiltonian eigenstates
with separate spins, and there is one to one correspondence between the S, and \To states to the $\uparrow\downarrow$ and $\downarrow\uparrow$ ones.
Conversely, for the potential variation in the opposite direction one obtains either the S or \To state, depending on the spin distribution
$\uparrow\downarrow$ or $\downarrow\uparrow$ in the left and right quantum dots respectively. This fact is next used in the detection of the spin exchange
with the spin and charge conversion induced by the potential variation. Obviously, the correspondence between S, \To and $\uparrow\downarrow$, $\downarrow\uparrow$ states
can be opposite with equal probability for a random HF field distribution,
however one or the other is typical for the HF field generated at random
(i.e. spins tend to separate at low $\Delta{}V$).

\subsection{Time evolution: Spin exchange}
Above we described how the system initialized  in the (0,2)S ground state 
is taken across the avoided crossing to the (1,1)S state and next into the separated spin state $\uparrow\downarrow$ or $\downarrow\uparrow$, depending on the state of the HF field. 
Here, we consider the rapid rise of the difference of potentials (the center of Fig. \ref{swap}) 
for which a nonzero exchange energy $J=E_{\mathrm{T}_0} - E_\mathrm{S}$ appears.
Then, the solution of the time dependent Schr\"odinger equation reads
\begin{equation} \Psi(t)=\frac{1}{\sqrt{2}}\exp(-\frac{iE_\mathrm{S}t}{\hbar})\left( \Psi_\mathrm{S}+\Psi_{\mathrm{T}_0}\exp(\frac{-iJt}{\hbar}) \right)\label{la} \end{equation}
and the spins in both the dots flip with the period of $T_f=\frac{2\pi\hbar}{J}$. The wave function of this form switches the spin orientations
within the dot, as it varies from $\Psi_{\uparrow\downarrow}$ (at $t=0$) to $\Psi_{\downarrow\uparrow}$ at $t=T/2$.
 The spin flips are stopped when the system is taken down to a small value of $\Delta V$ again. 
For spin exchange times which are odd multiples of $T_f$, a slow rising of the potential takes
the system to (1,1)\To eigenstates. For spin exchange times that are even multiples of $T_f$ the potential variation
returns the system to (1,1)S and next to (0,2)S. 

\begin{figure}[htbp]
	\includegraphics[width=3.4in]{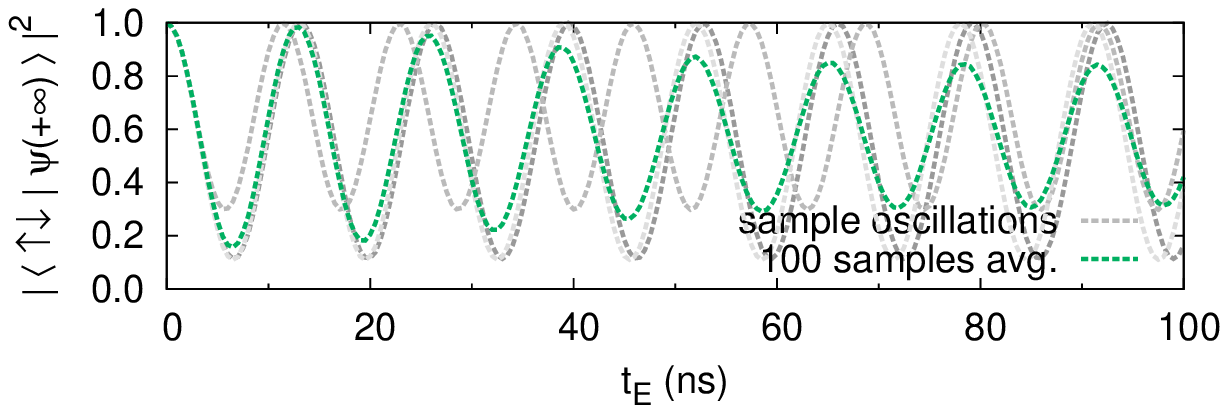}
	\caption{
	Probabilities $P_S = | \langle S | \psi \rangle |^2$ of finding the
	system in state S as a function of exchange time $t_E$, for which larger potential
	difference was reintroduced (to $\Delta{}V = 4.5$~meV). Gray plots are sample Rabi
	oscillations, while the green is an average over 100 random distributions of the HF field. \label{czcz}
	}
\end{figure}

\begin{figure}[htbp]
	\includegraphics[width=3.4in]{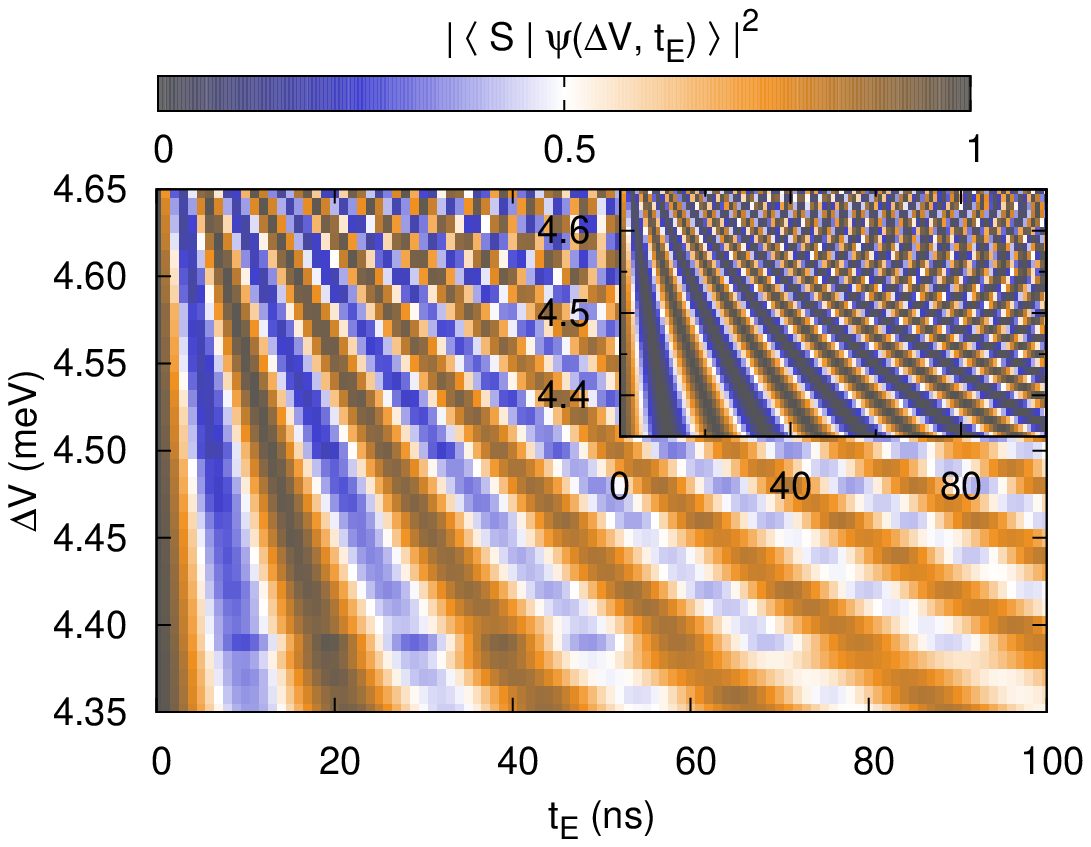}
		\caption{\label{1-swap} 
	Square of the absolute value of the projection of the wave function on the (1,1)S singlet state  as a function of exchange time $t_E$ and potential
	difference $\Delta{}V$. Every horizontal strip of the plot
	is an average of 60 simulations, each run for different HF sample.
	Inset presents the results expected for the time evolution given by Eq. \eqref{la} for the exchange energy determined by the $\Delta V$ [see Fig. 2(c)]. }
\end{figure}

Figure \ref{czcz} shows the projection of the wave function on $\Psi_{\uparrow\downarrow}$ -- taken as the initial state for the central high $\Delta V$ point in the time sequence as a function of time. The gray lines indicate three sample evolutions for some fixed random HF field orientations and the green one is
an average over 100 such runs. We can see that the average has a decreasing amplitude of the oscillations which corresponds to the
inhomogeneous broadening due to the random field. In Fig. \ref{1-swap} we show the projection of the wave function on the $\Psi_{\uparrow\downarrow}$  state as a function of the potential difference $\Delta{}V$ during the spin exchange and the spin exchange time $t_E$. The inset shows the result obtained by Eq. \eqref{la} for the
exchange energy as calculated in the absence of the HF field. The simulated pattern of the fringes agrees with the analytical one.
The result of the simulation contains the effect of the inhomogeneous potential -- the visibility of the oscillations deteriorates with the exchange time. The oscillations are closer to the ideal value for a large potential variation $\Delta V$ in which the interdot tunnel coupling and the exchange energy is larger. The spin exchange is not only faster but occurs with the larger fidelity for larger $\Delta V$, i.e. closer to the (1,1) -- (0,2) avoided crossing.

\section{Summary and Conclusions}

We presented the results of the simulation of the spin separation, exchange and spin-to-charge conversion for a model of 
a two-electron system confined in a double quantum dot defined within GaAs quantum wire
and a texture of the HF nuclear magnetic field with the SO interaction using the configuration interaction
method and an effective Overhauser field distribution.

For the potential difference (detuning) sweeps through the (0,2)S -- (1,1)S avoided crossing -- that is used for preparation
of the initial state as well as for the spin to charge conversion applied for the readout, 
the HF field and the SO interaction play similar roles, and one can be replaced by the other.
The transition times in HF field differ from one random distribution to the other, while
the SO interaction introduces an anisotropy of the evolution in the external magnetic field.
A simultaneous presence of the HF field and the SO interaction stabilize the variation range of the 
transition times and reduce the anisotropy as a function of the external magnetic  field orientation.

The spin-separation at the preparation stage is achieved due to the HF field and an adiabatic evolution
at the scale of the nuclear Zeeman effect. The rate of the spin exchange induced by a pulse of potential $\Delta{}V$
differs strongly from one random nuclear spin distribution to another, but the averaged spin evolution 
closely follow the time scale set by the exchange energy in the absence of the HF field.

\end{document}